# Constellation Design for Channels Affected by Phase Noise

Farbod Kayhan
Politecnico di Torino
Dipartimento di Elettronica e Telecomunicazioni
10129, Torino, Italy
Email:farbod.kayhan@polito.it

Guido Montorsi
Politecnico di Torino
Dipartimento di Elettronica e Telecomunicazioni
10129, Torino, Italy
Email: guido.montorsi@polito.it

*Abstract*—In this paper we optimize constellation sets to be used for channels affected by phase noise. The main objective is to maximize the achievable mutual information of the constellation under a given power constraint. The mutual information and pragmatic mutual information of a given constellation is calculated approximately assuming that both the channel and phase noise are white. Then a simulated annealing algorithm is used to jointly optimize the constellation and the binary labeling. The performance of optimized constellations is compared with conventional constellations showing considerable gains in all system scenarios.

## I. INTRODUCTION

This paper investigates the constellation design for transmission over AWGN channels affected by phase noise. Our main goal is to show that a significant improvement with respect to the conventional QAM and PSK constellations can be obtained by optimizing the constellation space considering both thermal and phase noise effects in the optimization process. The shape of the optimized constellation depends strongly on the communication system model, the channel model, the power constraints and the objective function to be optimized. Even for small signal sets with only 8 signals several different constellations can be obtained, each being optimal for a given set of parameters. We refer the readers to [1], [2], [3] and [4] and the reference within for a detailed study of the algorithms and methods used for constellation optimization over the AWGN channel.

Recently ( [5],[4]) we have proposed a joint signal-labeling optimization scheme for designing constellations which maximize the (pragmatic) achievable mutual information under the peak power constraint as a function of signal to noise ratio (SNR) using a simulated annealing (SA) algorithm for transmission over non linear satellite channels. In this paper we use the same algorithm focusing on the channels with phase noise, imposing an average power constraint.

In order to use the SA algorithm we first need to calculate the mutual information of the constellation for a channel affected by white additive Gaussian and phase noise. The problem of calculating the channel capacity impaired by additive Gaussian and phase noise is open in general and is an active research theme. Some asymptotic results for discrete-time channels at high SNR has been proposed by Lapidoth in [6] for phase noise both with and without memory. In [7] Kats and Shamai show that the capacity-achieving distribution of the discrete-time non coherent channel under the average power constraint is discrete with an infinite number of mass points. Also some upper and lower bounds on channel capacity are obtained. In particular, the upper bound is shown to be tight for asymptotically high SNR. More recently, Barbieri and Colavolpe have used a Monte Carlo simulation method to estimate the achievable information rate of the AWGN channel with phase noise, modelled as a Wiener process, assuming a uniformly distributed input [8] and [9]. However these results are either too complex computationally or valid only in the larger limit of constellation points. In [10] the authors calculate the mutual information for any given constellation with uniform distribution over the signal space and assuming the memoryless phase noise modelled as Tikhonov distribution. In this paper we provide a very close approximation for the mutual information which is computationally much faster.

Optimizing the constellation space under the phase noise has achieved a lot of attention in the literature. In [11] a 64-ary constellation is patented which is robust over the phase noise channels. In [12] the authors first provide an approximate expression for the symbol error rate of a given constellation in the presence of phase jitter and then use a gradient descent algorithm for finding a constellation which minimizes this expression. This approach does not take into the account the SNR as an optimization parameter. In the same paper the authors also explore the design of circular (APSK) constellations for use over phase noise channel. In [13] the authors optimize the 16-ary APSK constellations in the presence of nonlinear phase noise. In the first step the authors fix the number of rings and the number of signals over each ring. In the second step they find the optimal radii and phases of each ring minimizing the symbol error rate following the detectors and decoders suggested in [14].

Our approach in this paper is different from the previous works in several ways. First of all we provide an expression for the achievable mutual information (AMI) in the presence of white phase noise and thermal noise. The AMI is then maximized as a function of thermal and phase noise variances under a given power constraint. Furthermore, we do not impose any particular a-priori structure over the constellation space and our method allows for a joint signal-labeling optimization when the pragmatic achievable mutual information (PAMI) is the target function to be maximized [15].

The rest of this paper is organized as follows. In section II we describe the notations and formulate the optimization problem. An approximate fast method for calculating AMI and PAMI in the presence of phase noise is also presented in this section. We also provide a short description of the optimization algorithm. The optimization results for 8-ary constellations are presented in section III where we also provide several capacity curves of several optimized constellations. In section IV we provide end-to-end simulation results of our optimized constellations. Finally we conclude the paper in section V and discuss some future works.

## II. STATEMENT OF THE OPTIMIZATION PROBLEM

We consider a complex constellation $\chi$ with $M = 2^m$ elements. The elements of $\chi$ are referred as constellation points or simply signals. The Euclidean distance between two points in the complex plane is denoted by $d(.,.)$, and $d_H(.,.)$ is used to denote the Hamming distance between two binary sequences. The signals are associated to the bits at the input of the modulator through the one-to-one labeling $\mu : \chi \to \{0,1\}^m$. In particular, for any given signal $x$, $\mu^i(x)$ is the value of the $i^{th}$ bit of the label associated to it. A labeling for $\chi$ is called a Gray mapping if for any two signal $x_i, x_j \in \chi$ we have $d_H(\mu(x_i), \mu(x_j)) = 1$ if $d(x_i, x_j) \leq d(x_i, x_k)$, for all $x_k \in \chi$.

### A. Channel model and assumptions for the optimization problem

We consider a Complex Additive White Gaussian Noise (AWGN) channel with phase jitter, i.e., the received signal can be written as

$$y = xe^{j\varphi} + n$$

where $n$ is a complex Gaussian random variable with concentration $K_n = 1/\sigma^2$, $x$ is the transmitted constellation point and $\varphi$ is the phase rotation. We assume that $\varphi$ has a Tikhonov (also known as Von Mises) probability distribution function (PDF) with concentration $K_\varphi$, i.e.,

$$p(\varphi) = \frac{e^{K_\varphi \cos(\varphi)}}{2\pi I_0(K_\varphi)},$$

where $I_0(.)$ is the modified Bessel function of the first kind of order zero. As we have mentioned we consider only the residual phase error caused by the imperfect phase synchronization, and therefore the obtained results are independent from any specific phase-estimation technique. This simplification allows us to obtained a simple and fast approximation for the mutual information of any given constellation.

Our goal is to find the constellation that maximizes AMI of the channel, when all the constellation points are transmitted with the same probability. Therefore we wish to solve the optimization problem

$$\chi^* = \arg\max_\chi C(\chi) \tag{1}$$

where

$$\begin{aligned}C(\chi) &= E_{x,n,\varphi}\left\{\log_2 \frac{p(y|x)}{\frac{1}{M}\sum_u p(y|u)}\right\} \\ &= m - E_{n,x,\varphi}\left\{\max_u{}^* \log \frac{p(y|u)}{p(y|x)}\right\},\end{aligned} \tag{2}$$

where we used the short hand notation $\max^* \triangleq \log \int \exp$. Note that in general one needs to solve the above optimization problem under some power constraints over the constellation space. In this paper we focus only on average power constraint and fix it to one.

In order to calculate $C(\chi)$ we need first to compute $p(y|u)$ for any given input/ouput signals $u$ and $y = ue^{j\varphi} + n$. Since $\varphi$ is assumed to be a random variable with Tikhonov distribution we have

$$\begin{aligned}\log(p(y|u)) &= \log \int_{-\pi}^{\pi} p(y|u,\varphi')p(\varphi')d\varphi' \\ &= B + \max_{\varphi'}{}^*\left[\lambda(\varphi') - \frac{K_n}{2}|y - ue^{j\varphi'}|^2\right] \\ &= B - \frac{K_n}{2}|u|^2 + \max_{\varphi'}{}^*\left[\lambda(\varphi') + K_n\Re(y^* u e^{j\varphi'})\right],\end{aligned} \tag{3}$$

where we have introduced $\lambda(\varphi) = K_\varphi \cos(\varphi)$, and we have collected in the variable $B$ all constant terms w.r.t. $u$.

The main step to obtain a fast and accurate approximation of the mutual information is to substitute the $\max^*$ operation in last row of (3) with the maximum operation. Let $z \triangleq y^*u$ and $f(\varphi) = \lambda(\varphi) + K_n\Re(ze^{j\varphi})$, to find the maximum we need to find the solutions of the equation $\frac{\partial f}{\partial \varphi} = 0$ resulting in

$$\left.\frac{\partial \lambda(\varphi)}{\partial \varphi}\right|_{\varphi=\phi} = K_n \left.\frac{\partial}{\partial \varphi}\Re(ze^{j\varphi})\right|_{\varphi=\phi}$$

obtaining the following expression for $\phi$

$$\sin(\phi) = -|z|\frac{K_n}{K_\varphi}\sin(\arg z + \phi)), \tag{4}$$

where $\phi$ is a function of $z$. Equation (4) has the solution

$$\phi(z) = -\arctan\frac{A\Im(z)}{1 + A\Re(z)}, \tag{5}$$

where we have defined $A \triangleq \frac{K_n}{K_\varphi}$. Substituting the above equation in (3) we obtain

$$\log(p(y|u)) \approx B - \frac{K_n}{2}|u|^2 + \lambda(\phi(y^*u)) + K_n\Re(y^*ue^{j\phi(y^*u)}).$$

The normalized logarithm in (2) can then be approximated as

$$\log\left(\frac{p(y|u)}{p(y|x)}\right) \approx \lambda(\phi(y^*u)) - \lambda(\phi(y^*x))$$
$$+ \frac{K_n}{2}\left(2\Re(y^*ue^{j\phi(y^*u)}) - 2\Re(y^*xe^{j\phi(y^*x)}) - |u|^2 + |x|^2\right). \tag{6}$$

Finally, in order to calculate the mutual information we need to first calculate the $\max^*$ operation of (6) with respect to $u$ and then take the average with respect to both $n$ and $\varphi$ in (2). To rapidly compute these averages we used the Gauss-Hermite quadrature formula with degree $k = 7$ per each of the three noise dimension [16].

In the "pragmatic" receivers where the detector and the binary decoder are separated and no iterations take place between them the relevant measure to be optimized is the PAMI (see [15]). For a given constellation $\chi$ and labeling $\mu$, the symmetric PAMI of the channel is defined as

$$\mathcal{C}_P(\chi,\mu) = \sum_{i=1}^{m} I(\mu^i(X);Y) \leq C(\chi), \qquad (7)$$

where $X$ is the transmitted signal, $Y$ is the received symbol, and $I(.;.)$ denotes the mutual information function. For PAMI, the loss in terms of channel capacity compared to optimal joint detection and decoding, not only depends on the constellation, but also on the labeling. In general, non-Gray mappings induce a higher loss of capacity at high SNR's. The approximated computation of (7) can be performed using the previously described approach with some slight modifications that we omit for brevity (for detail the reader is referred to [4]).

### B. Simulated annealing algorithm for joint signal/labeling optimization

In [5] simulated annealing (SA) algorithm has been used for maximizing $\mathcal{C}(\chi)$ and $\mathcal{C}_P(\chi,\mu)$ in the absence of the phase noise. We use the same algorithm for optimizing the constellation for phase noise channels. SA, under some conditions on the cooling schedule, guarantees convergence to the global optimum. However, these conditions are not usually feasible, as they impose a very slow cooling schedule. When these conditions are not satisfied, one may argue about the optimality of the obtained constellations. Therefore it is important to choose carefully the parameters of the SA algorithm in order to provide a *good* local optimum.

The main adaptation needed to speed up the SA algorithm for maximizing the capacity is to define a maximum displacement length as a function of time. For details on such adaptation we refer the readers to [5] and the references within. The output of this algorithm -with the chosen parameters- is independent of the initial conditions for constellations with up to 16 signals, suggesting that the algorithm is optimal for such small constellations. However for larger constellations the SA algorithm becomes very slow and its output depends on the initial conditions.

### III. OPTIMIZATION RESULTS

In this section we present some of the 8 point optimized constellations and their corresponding capacity curves as a function of both SNR $\triangleq K_n/2$ and the phase noise standard deviation PNSD=$\sigma_\varphi = \sqrt{1/K_\varphi}$. We also compare the capacity curves of these optimized constellations with conventional 8-PSK constellation. As we will see, the shape of the optimized constellation strongly depends on both SNR and $\sigma_\varphi$. Note that for each pair of $\sigma_\varphi$ and SNR values and the chosen objective function (AMI or PAMI) one needs to run the SA algorithm to find the optimized constellation. In figure 1 we plot both AMI and PAMI curves of the optimized constellations for several values of PNSD and values of SNR from 1 to 15. The solid and dashed curves correspond respectively to AMI and PAMI of the optimized constellations. As it can be noticed, the loss due to the pragmatic approach is quite small (less than 0.2 dB) in all cases. This is true also for higher order constellations. In Figure 2 we present the optimized 8-ary constellations for six different values of PNSD with the SNR fixed into 12 dB, and obtained maximizing the AMI. For comparison, in Figure 3 we also show two constellations obtained by optimizing PAMI. As it can be seen, the constellations are slightly different from those in the previous figure.

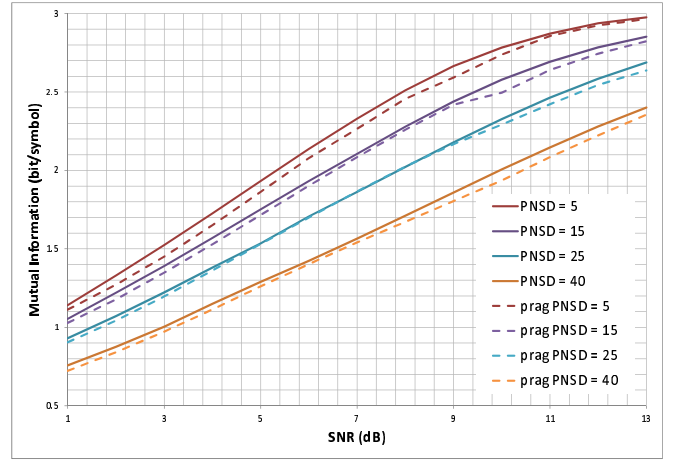

Fig. 1. AMI and PAMI for optimized constellation for various $\sigma_\varphi$ and SNR values.

To check the robustness of designed constellation to the mismatch with respect to the actual channel SNR in Figure 4 we report the AMI of the constellations of Figure 2, optimized for a target SNR of 12 dB, as a function of the actual channel SNR (solid lines). Comparing these curves with the ones in Figure 1 one can see that the constellations optimized at SNR=12 have performances close to the optimized constellations even for lower channel SNR values. However this is true only for small constellations. In Figure 4 we also plot the AMI curves for the constellation optimized at $\sigma_\varphi = 0$ (only AWGN) but used over the channel affected by phase noise (dashed lines). The loss due to the constellation suboptimality in this case is dramatic, showing that a constellation design that neglects the presence of phase noise can lead to very poor system performance. In particular, the loss increases with the target AMI to the point where it becomes infinite, as the AMI curves saturate below the maximum value of 3.

To check the robustness of designed constellation to the mismatch with respect to the actual channel PNSD in Figure 5 we plot the AMI of the constellations of Figure 2 as a function of the channel PNSD. We also plot for reference the capacity curve for the 8-PSK constellation. Constellations designed for a target PNSD does not degrade their performance robustness when the channel PNSD is smaller, while their performance rapidly degrades when it becomes larger. Finally, as it can be seen in Figure 5, the constellations optimized for high PNSD do not always saturate to the maximum spectral efficiency when used over channels with a lower phase noise standard deviation. Thus, there is a trade-off between the performance of the optimized constellations with respect to SNR and PNSD. Also we should notice that the optimized constellation over AWGN channel is not robust even for small channel PNSD values. This becomes particularly important when one needs to choose a constellation for a system where the PNSD is not accurately estimated, as it may cause a large loss in performance. These observations remain valid for larger constellations sets where the constellation performance becomes even more sensible to small changes of PNSD.

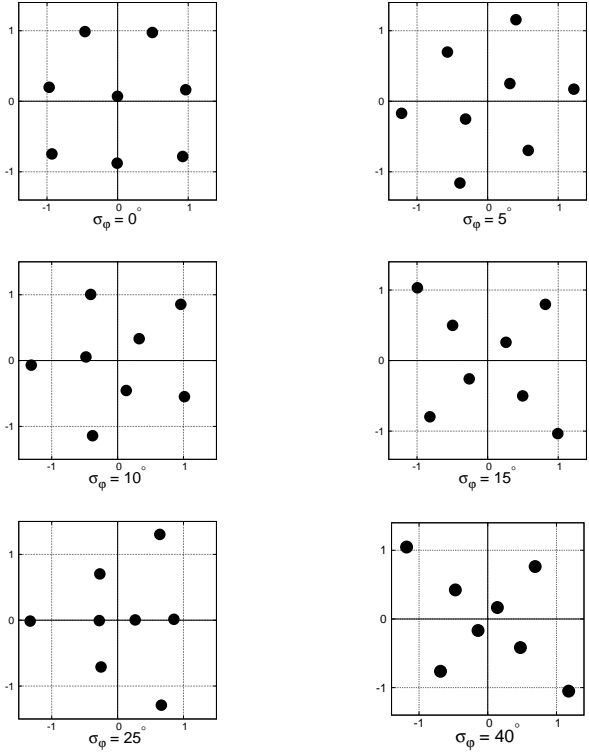

Fig. 2. Optimized 8-ary constellations for six different values of $\sigma_\varphi$. The objective function is the AMI and the SNR is fixed into 12 dB in all cases.

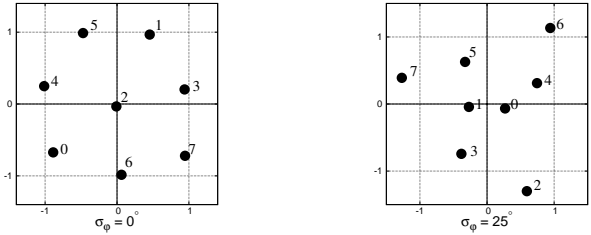

Fig. 3. Optimized 8-ary constellations for two different values of $\sigma_\varphi$. The objective function is the PAMI and the SNR is fixed to 12 dB in both cases.

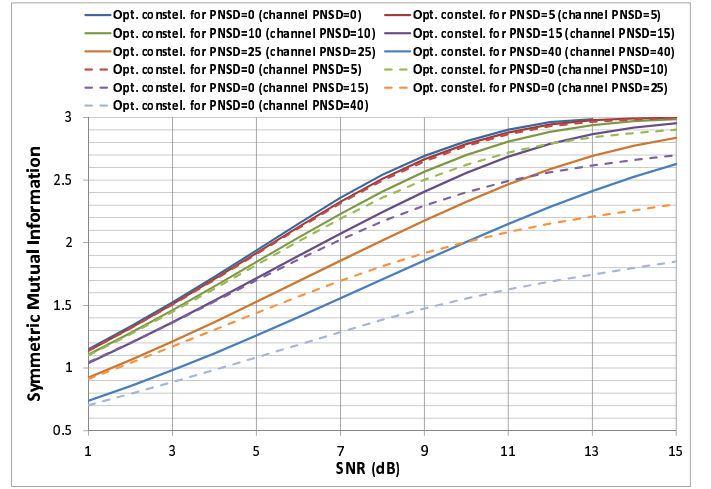

Fig. 4. Performance of optimized constellations in Figure 2 as a function of SNR. The solid lines are the AMI for the optimized constellations for the given PNSD, the dashed-line curves are the AMI for the constellation optimized for the AWGN but used in the presence of the phase noise.

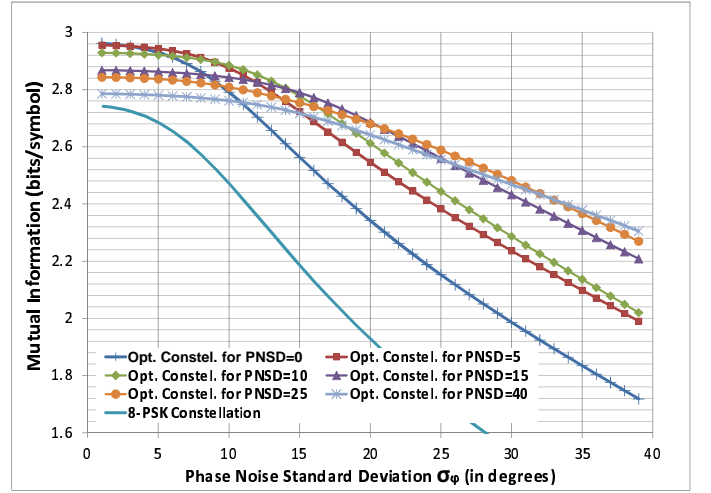

Fig. 5. Performance of optimized constellations in Figure 2 as a function of the phase noise standard deviation $\sigma_\varphi$.

## IV. SIMULATION RESULTS

In order to check the validity of our design approach, we have simulated the performance of the optimized constellations over a realistic system, using a powerful "capacity achieving" encoding scheme. In Figure 6 we report the results of the comparison of a set of 8 point constellations pragmatically encoded with a rate 5/6 binary serially concatenated convolutional code (SCCC). The details of the encoder structure can be found in [17]. We have considered the constellations reported in Figure 3, which have been optimized for the PAMI function fixing SNR=12 dB and PNSD=0 and 25 degree. Looking at Figure 4 one can verify that this SNR corresponds to a PAMI of 2.5, thus justifying our choice for the code rate ($R = 5/6$). Dashed curves refer to the AWGN channel without phase noise (channel PNSD=0). In this case the optimal constellation outperforms the 8-PSK constellation, reported for reference, by roughly 0.8 dB. The constellation designed for PNSD=25 shows a loss of about 0.3 dB w.r.t. 8-PSK. Solid lines refer to the case of AWGN channel affected by phase noise with PNSD=25. The situation in this case drastically changes. 8-PSK performance is not capable of delivering reliable communication for any value of SNR and its performance curve is not reported. The constellation optimized for PNSD=0 shows a loss of 19 dB, while the optimal constellation for PNSD=25 shows a loss of only 2.8 dB w.r.t. the AWGN channel and 3.9 dB w.r.t. the optimal constellation on the AWGN channel. Notice that these losses perfectly match those predicted by the PAMI computation in Figure 4 . Furthermore, to verify the difference between the PAMI and AMI optimization, in Figure 6 we also report the results obtained using a constellation optimized for the latter, associated to a Gray labeling. The constellation loses about 0.2 dB.

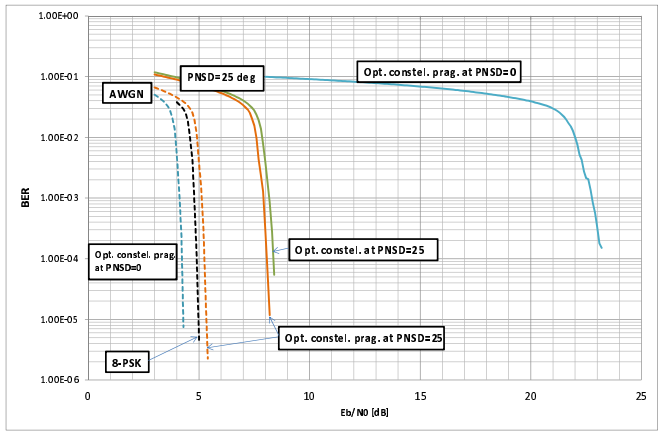

Fig. 6. Performance comparison of several 8 points constellations using a rate 5/6 SCCC code. Dashed lines refer to the performance over the AWGN channel, solid lines refer to the AWGN channel affected by phase noise with PNSD=25 degrees.

## V. CONCLUSIONS

In this paper we addressed the problem of constellation optimization for channels affected by phase noise. We first presented a fast approximated method to calculate the AMI of a given constellation for channels with white phase and thermal noise. Then the simulated annealing algorithm was used to maximize the AMI under the average power constraint. Our method allows also for a joint signal/labeling optimization of the constellation in pragmatic system scenarios, by maximizing the PAMI. We studied the performance of the optimized constellations both as a function of signal to noise ratio and of the phase noise standard deviation. We have also studied the robustness of our design to possible mismatches with respect to the actual channel conditions. In particular, it is observed that constellations optimized for AWGN channels alone are very sensitive to the phase noise, while other designs show more robustness to variations of PNSD and/or SNR. Finally to check the validity of our design approach we have simulated the optimized constellations over a realistic scenario including a capacity achieving code, obtaining results perfectly in line with the one predicted by AMI and PAMI calculations. This fact confirms that AMI and PAMI are the correct optimization functions to be considered for designing constellations to be embedded in system using turbo-codes. Even though we focused only on 8 point constellations, our method can be used to optimize much larger constellations. The results will be presented in an upcoming paper.

## REFERENCES

[1] A. J. Kearsley, "Global and local optimization algorithms for optimal signal set design," *J. Res. Natl. Inst. Stand. Technol.*, vol. 106, pp. 441–454, 2001.
[2] G. Foschini, R. Gitlin, and S. Weinstein, "On the selection of a two-dimensional signal constellation in the presence of phase jitter and gaussian noise," *Bell Syst. Tech. J.*, vol. 52, pp. 927–965, July-Aug. 1973.
[3] ——, "Optimization of two-dimensional signal constellations in the presence of gaussian noise," *IEEE Transactions on Communications,*, vol. 22, no. 1, pp. 28–38, Jan 1974.
[4] F. Kayhan and G. Montorsi, "Joint signal-labeling optimization under peak power constraint," *International Journal of Satellite Communications and Networking*, 2012.
[5] ——, "Joint signal-labeling optimization for pragmatic capacity under peak-power constraint," *Proceedings of IEEE Global Telecommunications Conference*, Dec. 2010.
[6] A. Lapidoth, "On phase noise channels at high snr," in *Information Theory Workshop, 2002. Proceedings of the 2002 IEEE*, oct. 2002, pp. 1 – 4.
[7] M. Katz and S. Shamai, "On the capacity-achieving distribution of the discrete-time noncoherent and partially coherent awgn channels," *Information Theory, IEEE Transactions on*, vol. 50, no. 10, pp. 2257 – 2270, oct. 2004.
[8] A. Barbieri and G. Colavolpe, "On the information rate and repeat-accumulate code design for phase noise channels," *Communications, IEEE Transactions on*, vol. 59, no. 12, pp. 3223 –3228, december 2011.
[9] L. Barletta, M. Magarini, and A. Spalvieri, "Estimate of information rates of discrete-time first-order markov phase noise channels," *Photonics Technology Letters, IEEE*, vol. 23, no. 21, pp. 1582 –1584, nov.1, 2011.
[10] T. Minowa, H. Ochiai, and H. Imai, "Phase-noise effects on turbo trellis-coded over m-ary coherent channels," *Communications, IEEE Transactions on*, vol. 52, no. 8, pp. 1333 – 1343, aug. 2004.
[11] S. Halyalkar, "64 qam signal constellation which is robust in the presence of phase noise and has decoding complexity," *U.S. Patent, no. 5832041*, Nov. 3 1998.
[12] Y. Li, S. Xu, and H. Yang, "Design of circular signal constellations in the presence of phase noise," in *Wireless Communications, Networking and Mobile Computing, 2008. WiCOM '08. 4th International Conference on*, oct. 2008, pp. 1 –8.
[13] L. Beygi, E. Agrell, and M. Karlsson, "Optimization of 16-point ring constellations in the presence of nonlinear phase noise," in *Optical Fiber Communication Conference and Exposition (OFC/NFOEC), 2011 and the National Fiber Optic Engineers Conference*, march 2011, pp. 1 –3.
[14] A. Lau and J. Kahn, "Signal design and detection in presence of nonlinear phase noise," *Lightwave Technology, Journal of*, vol. 25, no. 10, pp. 3008 –3016, oct. 2007.
[15] G. Caire, G. Taricco, and E. Biglieri, "Bit-interleaved coded modulation," *IEEE Transactions on Information Theory*, vol. 44, pp. 927–946, May 1998.
[16] R. Z. H. E. Salzer and R. Capuano, "Table of the zeros and weight factors of the first twenty hermite polynomials," *Journal of Research of the National Bureau of Standards*, vol. 48, no. 21, Feb 1952.
[17] S. Benedetto, R. Garello, G. Montorsi, C. Berrou, C. Douillard, D. Giancristofaro, A. Ginesi, L. Giugno, and M. Luise, "MHOMS: high-speed ACM modem for satellite applications," *Wireless Communications, IEEE*, vol. 12, no. 2, pp. 66–77, 2005.